\documentclass[aps,prl,twocolumn, groupedaddress, showpacs]{revtex4}
\usepackage{graphicx}
\begin{document}

\title{Probing the Upper Limit of Nonclassical Rotational Inertia}

\author{Ann Sophie C. Rittner}
\email{ar297@cornell.edu}
\affiliation{Laboratory of Atomic and Solid State Physics and the
Cornell Center for Materials Research, Cornell University, Ithaca,
New York 14853-2501}

\author{John D. Reppy}
\affiliation{Laboratory of Atomic and Solid State Physics and the
Cornell Center for Materials Research, Cornell University, Ithaca,
New York 14853-2501}

\date{\today}

\begin{abstract}
We study the effect of confinement on solid $^4$He's nonclassical
rotational inertia (NCRI) in a torsional oscillator by constraining
it to narrow annular cells of various widths. The NCRI exhibits a
broad maximum value of 20\% for annuli of ${\sim 100}$~$\mu$m width.
Samples constrained to porous media or to larger geometries both
have smaller NCRI, mostly below ${\sim  1}$\%. In addition, we
extend Kim and Chan's blocked annulus experiment to solid samples
with large supersolid fractions. Blocking the annulus suppresses the
nonclassical decoupling from 17.1\% below the limit of our detection
of 0.8\%. This result demonstrates the nonlocal nature of the
supersolid phenomena. At 20 mK, NCRI depends on velocity history
showing a closed hysteresis loop in different thin annular cells.
\end{abstract}


\maketitle

Kim and Chan (KC) have observed an anomalous decrease in solid
$^4$He's rotational inertia below 200~mK in a torsional oscillator
\cite{Kim2004,Kim2004a}. The possibility of a new ''super" state of
matter sparked a flurry of experimental and theoretical work. When
an annular cell is blocked, the nonclassical rotational inertia
(NCRI) is strongly reduced~\cite{Kim2004a}, indicating that
superflow is responsible for the NCRI. To date, the blocked-annulus
experiment is the strongest experimental evidence supporting
superflow over other explanations as  unusual temperature dependence
of the elastic properties of the solid~\cite{Kim2004a}. Further
support for superflow is that the oscillation frequency has no
impact on the signal size~\cite{Aoki2007}. The NCRI has been
confirmed in several laboratories~\cite{Kondo2006,Rittner2006} with
supersolid fractions ranging from 0.03\% up to 20\%
\cite{Rittner2007}. The supersolid fraction can be altered by
experimental parameters such as $^3$He impurity concentration
\cite{Kim2004}, thermal history of the sample~\cite{Rittner2006},
sample pressure, and geometric confinement~\cite{Rittner2007}.
Notably, the supersolid fraction increases by more than three orders
of magnitude when thin annular geometries confine the sample. From
experimental observations and a number of theoretical studies, there
has been a growing consensus in the field that crystalline defects
are crucial to at least enhance NCRIs (for a review
see~\cite{Balibar2008}). Some microscopic models suggest the
involvement of grain boundaries~\cite{Pollet2007}, dislocation
networks~\cite{Boninsegni2007}, a superglass phase
\cite{Boninsegni2006}, or a dislocation glass~\cite{Balatsky2006}.

The goals of our study are twofold: first, the sample confinement is
increased below 150~$\mu$m~\cite{Rittner2007} to maximize the
supersolid fraction. Second, we block annular cells with high NCRIs
to test if these samples also exhibit the characteristic superflow
behavior seen by KC.

We find that the supersolid fraction exhibits broad maximum of
around 20\% in narrow annuli of 100~$\mu$m. In such a cell, we
confirm the blocked annulus result~\cite{Kim2004}: inserting a block
in the flow path suppresses the supersolid fraction from 17.1\% to
below our experimental resolution of 0.8\%.

\begin{figure}[b]
\begin{center}
\begin{minipage}{8.6cm}
\vspace{-0.4cm}
\includegraphics[width=8.6cm]{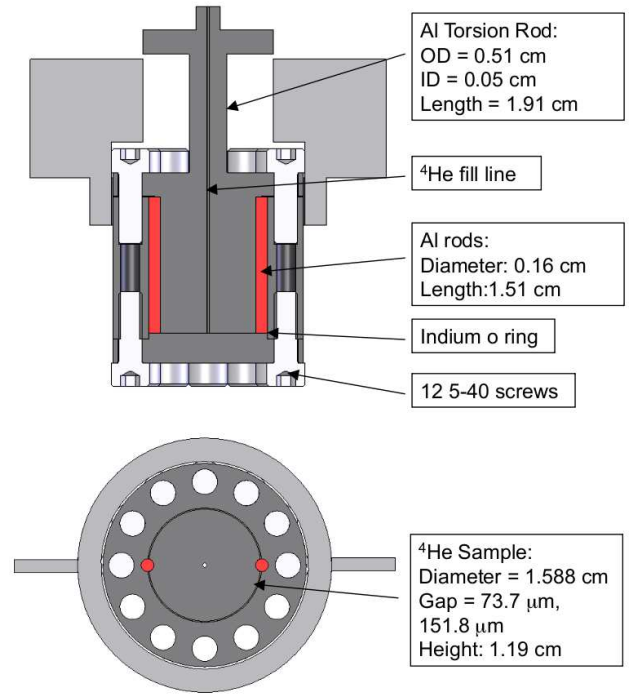}
\vspace{-.6cm} \caption{\label{setup}Aluminum torsional oscillator
with removable blocks (in red). The thin annular gap widths of
73.4~$\mu$m and 148.3~$\mu$m resulted in mass loading of 45.5~ns and
91.9~ns respectively. At 20~mK, the resonance frequency is 484.1~Hz
and the quality factor of the oscillation is $Q \sim 5.3 \times
10^5$.}
\end{minipage}
\end{center}
\end{figure}

In our latest design, we have constructed the torsion rod and the
body of the torsional oscillator from the aluminum alloy 6061T6.
Fig.\ \ref{setup} shows our aluminum torsional oscillator with an
annular geometry. Its resonance frequency is 484.1~Hz at temperature
$T$ = 4 K with a quality factor, $Q$, of 5.3$\times 10^5$. The inner
wall and torsion rod are made out of one piece to minimize the
relative motion of the two constraining walls of the annulus. This
design lessens the impact on the resonance period from $^4$He shear
modulus changes~\cite{Day2007}. Another unique feature of our
oscillator is that we can reversibly block it by introducing two
rods that are centered in the annulus (diameter~=~1.59~mm, shown in
red in Fig.\ \ref{setup}). This allows us to repeat KC's blocked
annulus experiment~\cite{Kim2004} in thin cells with large
supersolid fractions. The blocks also provide a mean to measure the
moment of inertia of the solid, which is needed to compute
supersolid fractions from the observed period drops. There are three
configurations for the oscillator: first the blocked configuration
with rods sealed in place; second,with slightly smaller diameter
rods to maintain an annulus of constant width; and third with the
rods absent, to study the effect on the NCRI of a larger region in
the path of the superflow. We employ annuli with two different
spacings, 73.4~$\mu$m, and 148.3~$\mu$m with surface to volume
ratios ($S/V$) of 134.8~cm$^{-1}$, and 272.5~cm$^{-1}$ respectively.

In most supersolid experiments, the total moment of inertia of the
solid is determined by the period increase upon freezing. In small
volume cells such as our narrow annuli, this increase is obscured by
a simultaneous decrease due to the dropping pressure. In our
experiment, a typical pressure drop of 30~bar in the cell during
solidification results in a period drop of 60~ns. For the
73.4~$\mu$m cell, this drop exceeds the 45.5~ns period rise from
solidification, making it impossible to use the standard
experimental method. Alternatively, we can determine the solid
inertia in our small volume cells by blocking the annulus: since the
fluid backflow is negligible in thin annuli, a block in the flow
locks the bulk liquid in the oscillator. When liquid enters the
cell, two effects cause the resonance period to increase: the
additional inertia stemming from the liquid $^4$He as well as the
cell's expansion due to the pressure. To separate pressure effects
from the period change due to coupling of the liquid, we measure the
resonance period as a function of liquid pressure in the cell. The
extrapolation of the period  to zero pressure is shifted with
respect to the zero pressure measured period before the cell was
filled. This period offset, $\Delta P$, is the period change that
stems from filling the cell with liquid at zero pressure. In order
to calculate the period shift due to solid helium $\Delta P$ is
rescaled by the ratio of solid to liquid density. The solid mass
loadings in the 73.4~$\mu$m and 148.3~$\mu$m cells are 45.5~ns, and
91.9~ns respectively. All supersolid fractions are calculated by
dividing the NCRI period drop by the solid mass loading.

We also use liquid $^3$He in calibrating our cell. Here, we take
advantage of the strong temperature dependence of the viscosity of
liquid $^3$He~\cite{parpia78}. Above 100~mK, the viscosity is low
and the fluid is mostly decoupled from the motion of the torsion
bob. As the temperature is lowered, the viscosity increases and the
fluid is increasingly locked in the annulus. The total fluid inertia
can be determined independently from temperature and height of the
dissipation maximum and from the period shift upon locking the
liquid. The mass loadings determined with both methods differ by
less than 5\%.

\begin{figure}
\begin{centering}
\hspace{-0.1cm}
\includegraphics[width=0.52\textwidth]{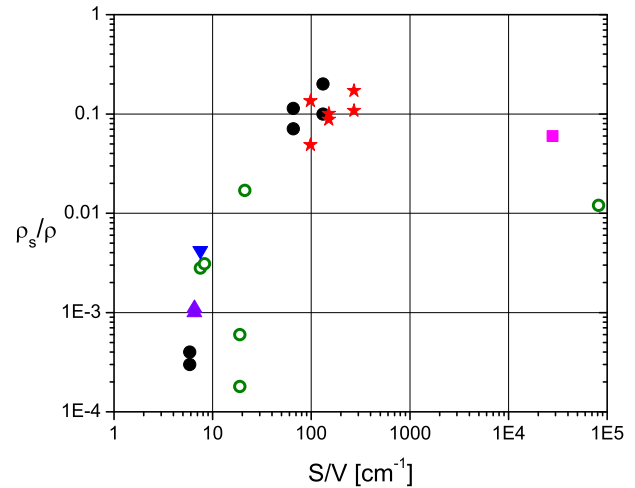}
\caption{\label{soverv} The supersolid fraction, $\rho_s/\rho$,
plotted as a function of the surface to volume ratio, which is
inversely proportional to the annular width. The geometries for the
different experiments from left to right are an open large cylinder
(solid circles)~\cite{Rittner2007}, a cylinder (triangles)
\cite{Aoki2007}, two slightly different cylinders (open circles)
\cite{Clark2007}, a cylinder (inverted triangle)~\cite{Kondo2006}, a
welded annulus (open circles)~\cite{Joshpriv}, an annular cell (open
circles)~\cite{Kim2004a}, thin annuli (solid circles and stars)
\cite{Rittner2007}, porous gold (square)~\cite{Kim2005}, and smaller
pore size porous gold (open circle)~\cite{Kim2005}. }
\end{centering}
\end{figure}

Fig.\ \ref{soverv} displays our main result of this series of
experiments; the supersolid fractions are shown as a function of
$S/V$ in different cells. In thin annular cells with gap, $t$, $S/V$
simplifies to $\frac{2}{t}$. For large open geometries, the
supersolid fraction is small, 0.03\%. As we have reported before,
the signal size increases dramatically by 3 orders of magnitude
\cite{Rittner2007} with stronger confinement. In our latest data
(solid stars), we find that the signals exhibit a broad maximum of
around 20\% at $S/V \sim 150$~cm$^{-1}$. When the sample is
constrained further, the NCRI decrease back to $\approx$ 1\% for
$S/V$ ${\sim 10^5}$~cm$^{-1}$. Quench-cooling or annealing fail to
alter the signal size.

Important information may be extracted from Fig.\ \ref{soverv},
particularly from the maximum NCRI and the length scale at which the
maximum occurs. First, the maximum NCRI rules out the simplest
explantation of superflow by a 3d network of grain boundaries. To
see this, we employ Kosterlitz-Thouless theory of thin films. It
relates the effective thickness of the grain boundaries and the
observed transition temperature, $T_c$, via
\begin{equation}
t = k_B T_c \frac{2 m^{* 2}}{\pi \hbar^2 \rho_s}
\end{equation}
where $k_B$ is the Boltzmann constant, $m^*$ is the effective mass
(here, the bare $^4$He mass), $\hbar$ the reduced Planck constant
and $\rho_s$ is assumed to be the bulk density. For experimentally
observed $T_c$'s, this gives a grain boundary thickness on the order
of tenths of Angstroms. Hence, obtaining the observed 20\% supersolid
fraction would require an unphysical grain sizes on the order of 1~$\AA$. Possibly, a less simple version of the grain boundary theory might be reconciled with our observations.

Similarly, the maximum NCRI is difficult to reconcile with a
dislocation network with superfluid cores. The measured dislocation
density in a sample space with $S/V$ = 2 cm$^{-1}$
\cite{Tsuruoka1979} is consistent with the expected supersolid
signal of 0.1\% in a similar geometry. On the other hand, a
supersolid fraction, $\rho_s$/$\rho$, of 20\% would require a
dislocation density of 10$^{13}$ cm$^{-2}$ assuming a superfluid
core of radius 6 $\AA$~\cite{Boninsegni2006}. This required density
is three orders of magnitude higher than the highest measured
dislocation density, corresponding to a spacing between dislocations
of 3~nm. Consequently, it is improbable that this simple model can fully
explain the supersolid results.

In a model that better accounts for the NCRI's geometry dependence,
disorder is concentrated in a layer close to the
walls~\cite{Hazzardprivate}. This picture is consistent with the
suggestion that dislocations form preferentially near cell walls in
solid helium \cite{Kosevich1983}. The surface roughness determines
the penetration depth of the dislocation network,
$\approx$~1-5~$\mu$m for a polished metal surface. The maximal NCRI
fraction is expected in an annular cell when the spacing is approximately twice
the disordered layer thickness. Assuming that the supersolid
fraction in the disordered region adjacent to the walls is 20\%, we
calculate the penetration depth to vary between 37~$\mu$m (current
data) and 169~$\mu$m~\cite{Rittner2007}. The penetration depth
varies less between cells within the same series than between cells
out of different materials. This larger variation may be related to
the differences in the surface roughness.

\begin{figure}
\begin{centering}
\includegraphics[width=0.51\textwidth]{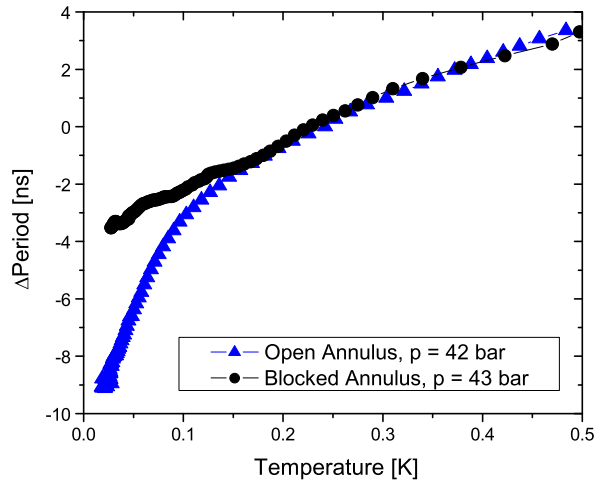}
\caption{\label{blockedann} Resonance period as a function of
temperature in an open (solid triangles) and blocked (solid circles)
annulus with a width of 73.4 $\mu$m. Both open and blocked annulus
data are taken in the same cell which could be reversibly blocked
(see Fig.\ \ref{setup}). The nonclassical rotational inertia
decoupling is 17.1\% of the solid inertia in the open annulus. For
the blocked annulus, the upper limit to the NCRI is 0.8\%
corresponding to a more than twentyfold reduction upon blocking. }
\end{centering}
\end{figure}

The second goal of our experiments is to check if the high apparent
supersolid fractions can still be attributed to long range
superflow. For this reason, we repeat KC's blocked annulus
experiment~\cite{Kim2004} in our narrow annular cells. In the
blocked annulus experiment, a partition is placed across the annular
channel, thus interrupting any long range flow around the annulus.
The basic idea is to compare the magnitude of supersolid signals in
an open and in a blocked annulus of the same width. In the
experiments performed to date, the solid helium moment of inertia
decreases upon blocking, indicating that the macroscopically
coherent supercurrent is suppressed. As a minor caveat, there
remains a small contribution to the NCRI from potential flow induced
by the rotational motion of the oscillator. In the limit of a long
narrow blocked channel, the NCRI from this backflow becomes
negligibly small compared to the NCRI for the unimpeded flow in an
unblocked channel. For example, in a 0.65~mm annulus, the period
drop is reduced 200-fold~\cite{Kim2004}. Our annular width is
73.4~$\mu$m, more than an order of magnitude smaller than Kim and
Chan's original cells (KC: open annulus gap = 0.95~mm, blocked
annulus gap = 1.1~mm \cite{Clark2008}) and the expected blocked
annulus backflow NCRI is below our resolution.

Fig.\ \ref{blockedann} displays the resonance period as a function
of temperature for both open and blocked annuli.  Upon blocking the
period drop at the supersolid transition is suppressed. Given the
noise level of the experiment, the upper limit on a residual period
drop in the blocked cell is $\sim 0.8$\%. The open cell displays a
supersolid fraction of 17.1\%, so the block suppresses the signal
more than twentyfold. The major advantage of our setup with regard
to KC's~\cite{Kim2004} is that our cell can be reversibly blocked,
allowing one to measure open and blocked annuli within the same
cell. We have also blocked a bigger annulus with a width of
487~$\mu$m and find the upper limit for a remnant supersolid signal
to be 0.4\%. In the open geometry, we expect the NCRI to be $\sim
5$\% (see Fig.\ \ref{soverv}).

Confirmation of the blocked annulus result demonstrates the nonlocal
nature of the supersolid phenomenon. Thus, local models, such as the
two-level tunneling systems~\cite{Andreev2007}, are unlikely to
provide a full explanation of the supersolid.

We have also studied the effect of removing the rods and thus
interposing a larger volume in series with the superflow. Although
the fractional signal is smaller, $\sim$~6\%, the total mass current
is not appreciably reduced.

\begin{figure}
\begin{centering}
\includegraphics[width=0.51\textwidth]{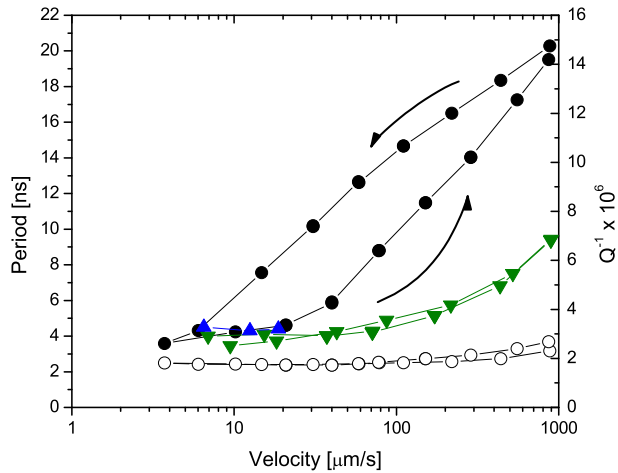}
\caption{\label{hysteresis} Velocity dependence of resonance period
(solid circles) and dissipation (open circles) at 20 mK in annular
cell with a width of 148.3~$\mu$m. First, the sample is cooled at a
high velocity, $v = 881$~$\mu$m/s, to 20 mK and then the drive is
reduced at constant temperature. After reaching a low velocity, $v =
3.8$~$\mu$m/s, the drive is increased again. The critical velocity
in this cell is $\sim $20$~\mu$m/s, as indicated by the low
temperature period of cool downs at a constant drive level
(triangles). We only show data points that were taken after the
oscillator had time to equilibrate, about 20~minutes. Arrows indicate the direction of
the velocity changes. For comparison, period data of the empty cell
(inverted triangles) are displayed.}
\end{centering}
\end{figure}

Finally, we have measured the velocity dependence of the supersolid
fraction below 40~mK~\cite{Aoki2007, Clark2008} in an annular cell
with a 148.3~$\mu$m gap. Fig.\ \ref{hysteresis} displays resonance
period (solid circles) and dissipation (open circles) as a function
of rim velocity, $v$, at 20~mK. The empty cell background (inverted
triangles) is displayed for comparison. Following a similar
experimental procedure as~\cite{Aoki2007}, we cool the sample to
20~mK while oscillating at a high rim velocity, $v=881$ $\mu$m/s.
Holding the temperature fixed, the velocity is decreased in steps
and then held for $\sim$~20~minutes until the amplitude came into equilibrium as determined by the oscillator's $Q$. Starting from the lowest velocity $v$ = 3.8~$\mu$m/s,
we raise the velocity in steps. When the velocity surpasses ${\sim
20}$~$\mu$m/s, the period rises more steeply than the empty cell
period, indicating that this cell's critical velocity has been
exceeded. The period difference between cell filled with solid
helium and empty cell at the highest velocity corresponds to a
supersolid fraction of 12.0\%. We observe some hysteresis between
decreasing and increasing velocity, that is, the resonance period
depends on the velocity history. In contrast, the resonance period
shows no hysteresis at 60 and 200~mK.

Our finding differs from Aoki \textit{et al.}'s observations in a
cylindrical cell~\cite{Aoki2007}. When their sample velocity
increases, the NCRI stays constant above the critical velocity of
15~$\mu$m/s, up to 800 $\mu$m/s. Also in a cylindrical cell, Clark
\textit{et al.}~\cite{Clark2008} find a correlation between the
sample growth method and the NCRI stability when the velocity is
increased: constant pressure grown samples with relatively low NCRI
are metastable at low temperatures, while the NCRI of blocked
capillary grown samples is unstable against an increase in velocity.
They attribute the existence of metastable states to severe vortex
pinning in the sample at low temperatures, in qualitative agreement
with Anderson's vortex liquid model~\cite{Anderson2007}. The major
difference between our experiments and other groups' lie in the
annular geometry, stronger confinement and much higher supersolid
fractions. Possibly, the smaller hysteresis in confined geometries
can be attributed to the fact that vortices cross the sample more
easily, for example because of a lower density of pinning centers.

\begin{acknowledgments}
We thank M.H.W. Chan, J.T. West, A.C. Clark, X. Lin, and E.J.
Mueller for providing us with information about their torsional
oscillators and extensive discussions, and we thank K.R.A. Hazzard
for a critical reading of the manuscript. This work has been
supported by Cornell University, the National Science Foundation
under Grant DMR-060584 and through the Cornell Center for Materials
Research under Grant DMR-0520404.
\end{acknowledgments}

\end{document}